\documentclass[letterpaper, 10 pt, conference]{ieeeconf}  

\IEEEoverridecommandlockouts                              

\title{\LARGE \bf
Evaluating Multi-Sensor Placement and Neural Network Architectures for Physical Activity Level Classification
}

\author{Bo Cui$^{1}$, Xiaowen Song$^{1}$, Tabak Monique$^{1}$, Bert-Jan van  Beijnum$^{1}$ and Ying Wang$^{1}$
\thanks{*This publication is part of the project LoaD (projectnr. NWA1389.20.009) of the NWA-ORC research programme which is (partly) financed by the Dutch Research Council (NWO)}
\thanks{$^{1}$Bo Cui, Xiaowen Song, Tabak Monique, Bert-Jan van  Beijnum, Ying Wang are with Faculty of Biomedical Signals and Systems, Electrical Engineering, Mathematics and Computer Science,
        University of Twente, 7500 AE Enschede, The Netherlands
        {\tt\small m.r.cui@utwente.nl}}%
}

\begin{document}

\maketitle
\thispagestyle{empty}
\pagestyle{empty}

\begin{abstract}

Accurate physical activity level (PAL) classification could be beneficial  for osteoarthritis (OA) management. This study examines the impact of sensor placement and deep learning models on AL classification using Metabolic Equivalent of Task values. The results show that the addition of an ankle sensor (WA) significantly improves the classification of high intensity activities compared to wrist-only configuration (53\% to 86.2\%). The CNN-LSTM model achieves the highest accuracy (95.09\%). Statistical analysis confirms multi-sensor setups outperform single-sensor configurations ($p<0.05$). The WA configuration offers a balance between usability and accuracy, making it a cost-effective solution for AL monitoring, particularly in OA management.

\end{abstract}

\section{INTRODUCTION}
Knee osteoarthritis (OA) is a degenerative joint disorder that significantly affects mobility, quality of life, and often leads to chronic pain and functional limitations \cite{c1}. One critical factor influencing both the progression and management of OA is physical activity level (AL), which encompasses the intensity and duration of daily movements \cite{c7}. Research indicates that moderate and consistent physical activity helps maintain joint function, alleviates pain, and improves overall health outcomes for OA patients \cite{c1}.  

The relationship between activity level and OA-related pain is complex. Burrows et al. \cite{c2} demonstrated that physical activity exerts both short-term and long-term effects on pain. In the short term, sudden increases in activity intensity can temporarily worsen pain, whereas sustained physical activity over time is linked to reduced pain, underscoring its potential benefits in OA symptom management. Among different activity levels, moderate-to-vigorous physical activity is particularly effective in pain relief, as it has a stronger negative correlation with pain than general movement. This highlights the importance of distinguishing between activity intensities, as overexertion may exacerbate symptoms, necessitating a balanced approach to activity duration and intensity. Accordingly, continuous monitoring of AL plays a crucial role in personalized OA pain management. 

The validated Metabolic Equivalent of Task (MET) can be seen as an accurate indicator of activity intensity and sedentary behavior, particularly in older adult\cite{c9}, has been known as a standard method to indicate the intensity of AL. The Compendium of Physical Activities \cite{c4} refines this classification by assigning MET values to various activities, enabling precise energy cost estimations across diverse contexts. 

Despite the widespread use of MET for PAL classification, current approaches focusing on conventional machine learning(ML) models. Reiss et al. \cite{c5} for example applied several conventional ML models to classify AL based on MET values, but few studies have explored similar methodologies, particularly leveraging deep learning techniques. To address this gap, we investigate the potential of deep learning for AL classification using MET. Additionally, in the field of human activity recognition, Nweke et al. \cite{c8} demonstrated that sensor placement significantly impacts classification outcomes, suggesting that different sensor configurations could influence classification accuracy.  Building on this finding, we hypothesize that sensor placement will similarly impact MET classification accuracy. To validate this, we aim to identify the optimal combination of inertial measurement unit (IMU) sensor and deep learning models for accurate AL classification. Specifically, this study has two key objectives:  

\begin{itemize}
    \item Compare various deep learning architectures for AL classification, focusing on distinguishing activity intensity (low, medium, and high) based on MET values.  
    \item Identify the optimal combination of wearable sensor placements (wrist, chest, and ankle) to enhance the accuracy and stability of AL classification.  
\end{itemize}

\section{Methodology}

\subsection{Dataset}
The PAMAP2 dataset\cite{c5} includes data collected from nine subjects, consisting of one female and eight males, with an average age of 27.22 years (±3.31) and a Body Mass Index (BMI) of 25.11 kg/m² (±2.62). The data was captured using three Colibri wireless IMUs and one heart rate (HR) monitor. These IMUs, which contain 3-axis MEMS sensors (accelerometers, gyroscope), were placed at three different body positions: the wrist, the chest, and the dominant ankle. The IMUs were sampled at 100Hz to capture comprehensive movement data. The dataset includes data from 12 predefined activities—such as lying, sitting, standing, walking, running, cycling, Nordic walking, ironing, vacuuming, rope jumping, and ascending/descending stairs—as well as a few optional activities, such as watching TV, computer work, driving, folding laundry, cleaning the house, and playing soccer. Over 10 hours of data were collected in total. Each activity's MET was also recorded to indicate activity intensity.

In this study, activities in PAMAP2 were categorized into three intensity levels—low, medium, and high—based on their MET values, adhering to international MET standards\cite{c4}. Low-intensity activities ($\leq 3$ METs) include sedentary behaviors such as lying, sitting, and watching TV. Medium-intensity activities (3-6 METs) encompass light to moderate exercises like walking, cycling, and household chores. High-intensity activities ($> 6$ METs) involve vigorous exercises such as running, rope jumping, and playing soccer. 

\subsection{Sensor Placement Configurations}

To systematically examine the role of different sensor configurations, we considered five distinct setups: 
\begin{enumerate}
    \item wrist with only an accelerometer (WO): Given the setup of the LoaD project’s two-year cohort study, where data was collected using a smartwatch equipped with only a three-axis accelerometer, the WO configuration naturally served as a baseline for AL classification based on simple motion detection. To establish an initial performance benchmark, we first evaluated the WO configuration using accelerometer-only data to determine its effectiveness in classifying AL.
    
    \item wrist with a full IMU (W6): With advancements in wearable technology, many modern smartwatches now incorporate six-axis IMUs, integrating both accelerometer and gyroscope data. In this study, we leveraged both accelerometer-only and six-axis IMU data to investigate whether incorporating rotational information from the gyroscope enhances classification accuracy.

    \item wrist and chest (WC): The chest serves as a stable reference point for monitoring upper-body movements. By integrating an additional chest-mounted IMU, we aimed to capture core body movement and posture-related variations that may not be fully detectable from the wrist alone. 

    \item wrist and ankle (WA): The ankle is a critical joint in human movement, particularly in weight-bearing activities such as walking and running. This configuration was chosen to determine whether supplementing wrist data with lower-body kinematics improves classification performance.

    \item wrist, chest, and ankle (W18): A multi-sensor setup incorporating wrist, chest, and ankle IMUs offers the most comprehensive representation of whole-body movement.By fusing data from all three locations, we aimed to assess whether a full-body sensing approach enhances classification accuracy beyond that of individual or dual-sensor setups.
\end{enumerate}

\subsection{Model}

In this study, five neural network (NN) architectures were explored for classifying AL from five configuration IMU data: Multilayer Perceptron (MLP), ResNet18, ResNet1D, Convolutional Neural Network, and Hybrid Conv1D-LSTM. Each model was designed and implemented to process the IMU time series data, leveraging unique architectural characteristics to improve performance on the task.

\subsubsection{Multilayer Perceptron (MLP)}

The MLP model served as a baseline, consisting of fully connected layers interspersed with activation functions. The straightforward architecture of MLP is effective for modeling non-linear relationships but lacks mechanisms to capture temporal dependencies in the data.

\subsubsection{ResNet18}

ResNet18, a variation of the ResNet family \cite{c6}, is known for its residual learning framework, which helps mitigate the vanishing gradient problem in deep networks. It is a residual network using skip connections to ease gradient flow. Its hierarchical structure extracts complex spatial features from IMU sequences.

\subsubsection{ResNet1D}

ResNet1D is an 1D adaptation of ResNet with fewer layers than Resnet18, tailored for time-series data. It combines residual blocks and global average pooling for efficient feature extraction.

\subsubsection{Convolutional Neural Network (CNN)}

The CNN model focuses on extracting spatial features from sequential IMU data using stacked 1D convolutional layers. These layers are designed to capture hierarchical spatial features from the input sequence. This architecture is computationally efficient and leverages local dependencies in the time-series data.

\subsubsection{Hybrid Conv1D-LSTM}

The hybrid Conv1D-LSTM model combines the feature extraction power of 1D convolutional layers with the temporal modeling capability of LSTM layers. The initial convolutional layers capture local spatial patterns from the IMU data, while the subsequent LSTM layers model long-term dependencies across the sequence. This hybrid approach enables the model to leverage both spatial and temporal features, making it particularly effective for time-series data such as IMU signals.
\subsection{Model Parameters and Hyperparameters}

In this study, several model architectures were evaluated. To ensure that the comparison of these models was consistent and meaningful, the hyperparameters were held constant across all architectures. This uniformity in hyperparameter settings allowed for a fair and direct comparison of the architectures' performance.

The key hyperparameters considered in this study included the learning rate, number of epochs, batch size, and regularization parameters. The learning rate was set to 0.01 as a fixed value across all models to provide stability during training and avoid large fluctuations in the optimization process. The number of epochs was set to 15 which was selected to ensure that each model had sufficient opportunity to converge, balancing between underfitting and overfitting. The batch size was set to 10 which was standardized to optimize both memory utilization and training efficiency across the different architectures. 

These hyperparameters were selected based on preliminary experiments, in which various values were tested on a subset of the dataset to determine an optimal configuration that balanced model performance and computational efficiency. This approach maintained uniform experimental conditions, thus isolating the impact of architectural variations on model performance. ReLU was set as the activation function in each model. 

\subsection{Model Training and Optimization}

The training process was structured in a way to ensure robust model evaluation and minimize biases in the results. The dataset was split into training and testing sets using an 80/20 random partition, ensuring that the models were trained on a diverse range of data and validated on unseen examples. This random splitting method reduces selection bias and supports the generalizability of the models to new data. Additionally, a Leave-One-Out validation strategy was applied during training to provide a more robust evaluation and reduce the risk of overfitting.  Optimization was carried out using stochastic gradient descent (SGD) with momentum.

The combination of these training and optimization strategies, along with the standardized hyperparameters, ensures that any differences in model performance can be attributed primarily to the architectural variations, rather than discrepancies in the experimental setup.

\subsection{Model Evaluation Metrics}
To evaluate the performance of the models, we used two primary metrics: accuracy and F1-score. Accuracy measures the overall correctness of the model's predictions by calculating the ratio of correctly classified samples to the total number of samples. F1-score, on the other hand, provides a balanced measure of precision and recall, especially useful for imbalanced datasets. These metrics were calculated for each sensor configuration and neural network model to assess their ability to classify activity levels effectively. Additionally, confusion matrices were generated to provide a detailed view of the classification performance for each activity level. These matrices allowed us to analyze the distribution of misclassifications and identify specific challenges in differentiating between activity classes. However, in the results of this paper, we only present confusion matrices for the WO and WA configurations. This selection was made to illustrate the baseline performance (WO) and the configuration that most suitable for daily monitoring (WA).

\subsection{Statistical Analysis}

To evaluate whether different sensor placements resulted in significant differences in classification performance, we conducted statistical analysis using the Wilcoxon signed-rank test on the F1 score. This non-parametric test was chosen due to its suitability for comparing paired data distributions, particularly when the sample size is small. 

To ensure the robustness of our analysis and mitigate concerns regarding the limited data size, we performed repeated tests on each participant's data. Specifically, for each participant, we conducted two independent evaluations on their test set to account for potential variability in the results. These repeated tests allowed us to better capture the inherent variability and ensure more reliable conclusions.

In addition, to adjust for multiple comparisons, we applied Bonferroni correction. Given that three pairwise comparisons were made (WO-WA, WO-W18, and WA-W18), the significance threshold was adjusted to \( \alpha = \frac{0.05}{3} = 0.0167 \). This correction ensured that the probability of Type I error was controlled across all comparisons.

\section{Result}

The results for both accuracy and F1-score are summarized in Tables \ref{tab:accuracy} and \ref{tab:F1score}, respectively.

\subsubsection{Accuracy}
Among the models, CNN-LSTM achieved the highest accuracy of 87.94\% in WO configuration, closely followed by Resnet-18, which demonstrated an accuracy of 87.33\%. Both models performed better when utilizing the W18 configuration, where accuracy peaked at 95.09\% and 95.49\%, respectively. The trend of increasing accuracy with the inclusion of more sensors was consistent across all models.

\subsubsection{F1-Score}
As presented in Table \ref{tab:F1score}, the CNN-LSTM model achieved the highest F1-score of 0.82 in WO configuration, followed by Resnet-18 with an average F1-score of 0.79. Again, the W18 configuration yielded the highest F1-scores for both models, reaching 0.94. This indicates that leveraging all available sensor data enhances the model's ability to classify correctly.
\begin{table}[h!]
\centering
\caption{Accuracy of all the networks}
\label{tab:accuracy}
\resizebox{1\linewidth}{!}{ 
\begin{tabular}{|l|c|c|c|c|c|c|}
\hline
\diagbox{Model}{Metric} & WO & W6    & WA    & WC    & W18  \\ \hline
MLP        & 67.39\%    & 73.64\% & 83.94\% & 81.78\% & 85.27\%  \\ \hline
CNN        & 88.24\%    & 84.97\% & 91.15\% & 86.67\% & 94.00\%  \\ \hline
CNN-LSTM   & 87.94\%    & 91.76\% & 94.3\% & 94.07\% & 95.09\%  \\ \hline
Resnet     & 87.45\%    & 90.18\% & 92.12\% & 93.85\% & 93.52\%  \\ \hline
Resnet-18  & 87.33\%    & 92.12\% & 94.79\% & 93.11\% & 95.49\%  \\ \hline
\end{tabular}}
\end{table}

\begin{table}[h!]
\centering
\caption{F1-score of all the networks}
\label{tab:F1score}\resizebox{1\linewidth}{!}{ 
\begin{tabular}{|l|c|c|c|c|c|c|}
\hline
\diagbox{Model}{Metric} & WO & W6    & WA    & WC    & W18     \\ \hline
MLP        & 0,63    & 0,71 & 0,81 & 0,77 & 0,83  \\ \hline
CNN        & 0,82    & 0,80 & 0,90 & 0,89 & 0,93  \\ \hline
CNN-LSTM   & 0,82    & 0,90 & 0,94 & 0,93 & 0,94  \\ \hline
Resnet     & 0,80    & 0,85 & 0,92 & 0,92 & 0,93  \\ \hline
Resnet-18  & 0,79    & 0,90 & 0,94 & 0,92 & 0,94  \\ \hline
\end{tabular}}
\end{table}

\begin{figure}[htbp]
    \centering
    \includegraphics[width=0.9\linewidth]{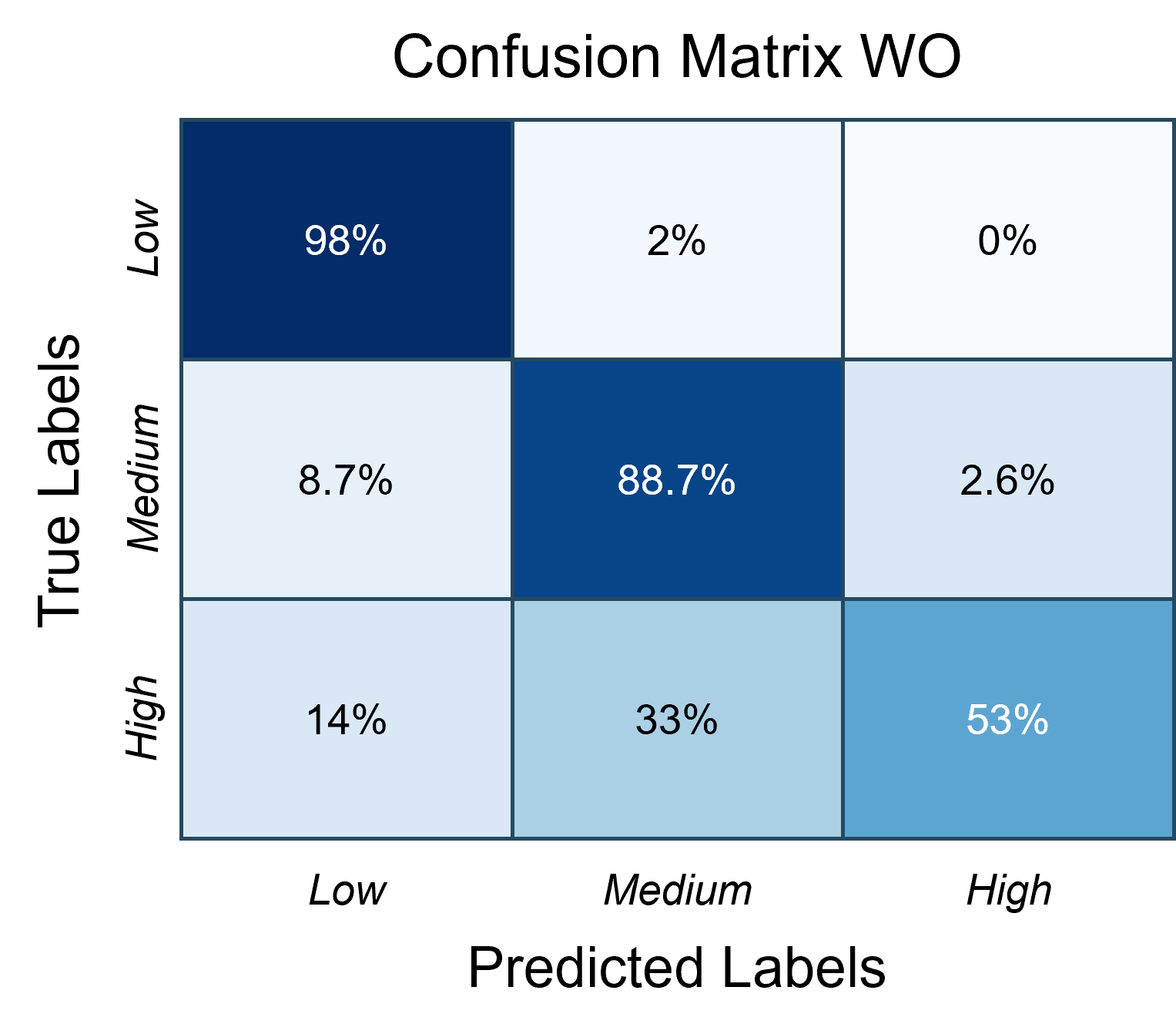} 
    \caption{Confusion matrix for the WO, presented in percentage.}
    \label{fig:image1}
\end{figure}

\begin{figure}[htbp]
    \centering
    \includegraphics[width=0.9\linewidth]{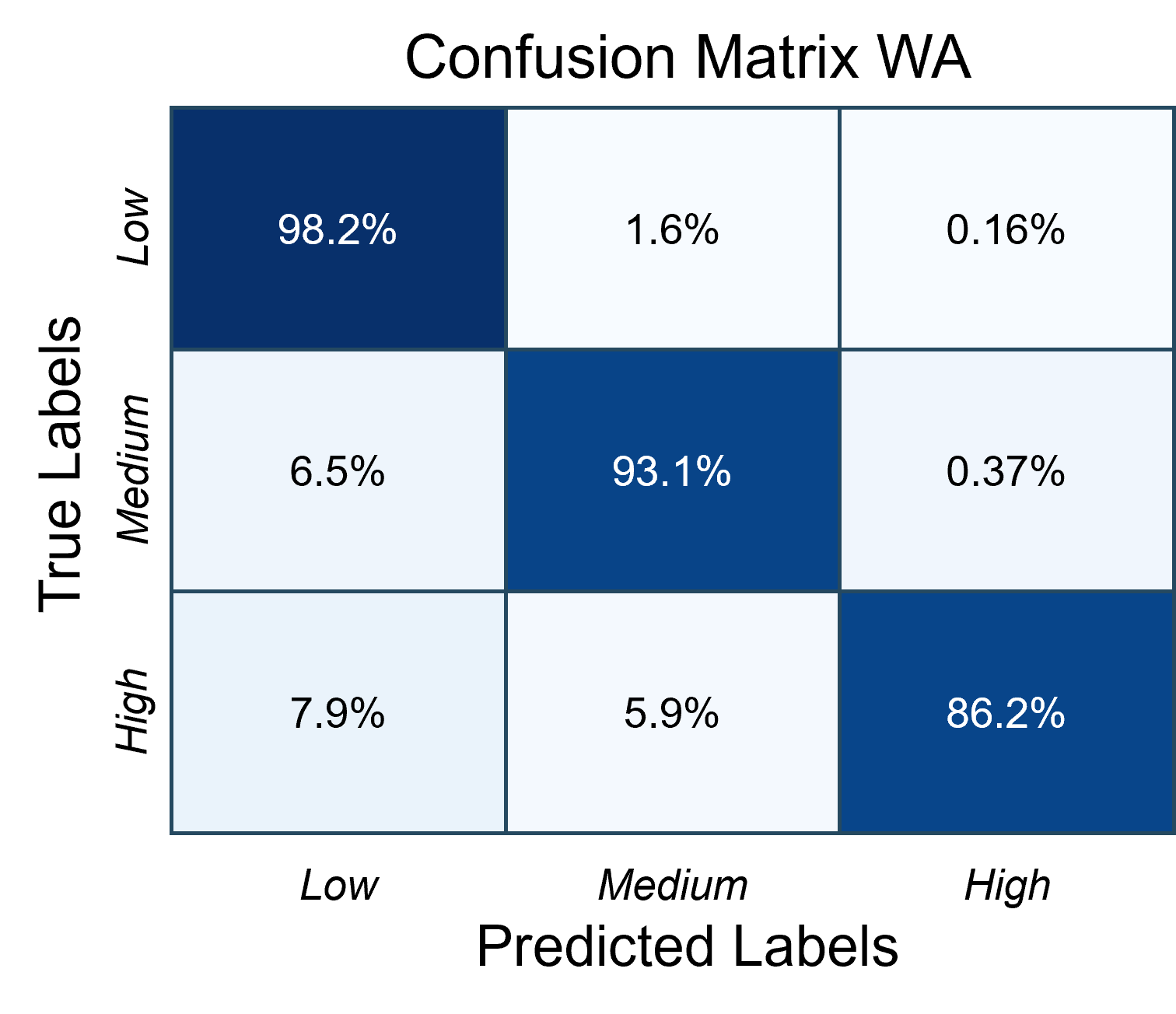} 
    \caption{Confusion matrix for the WA, presented in percentage.}
    \label{fig:image2}
\end{figure}

\subsubsection{Prediction Performance Across Sensor Configurations}

Figure \ref{fig:image1} and Figure \ref{fig:image2} present the confusion matrices for the WO and WA configurations, respectively. 

The WO configuration demonstrated excellent performance for low activity, achieving 98\% accuracy, and good performance for medium activity at 88.7\%. However, it struggled with high activity classification, reaching only 53\% accuracy, with 33\% of high activity instances misclassified as medium and 14\% as low. In contrast, the WA configuration significantly improved classification performance while maintaining similar accuracy for low activity (98\%). Medium activity classification improved from 88.7\% in WO to 93.1\% in WA, and high activity classification increased substantially from 53\% in WO to 86.2\% in WA. Misclassifications in WA were mainly between medium and high activity, with 7.9\% of high activity instances misclassified as low and 5.9\% as medium. These results indicate that adding sensors at multiple body locations (wrist and ankle) effectively captures complex motion patterns, enhancing classification accuracy, particularly for high-intensity activities.

\subsubsection{Statistical Comparison of Sensor Placements}

The comparison between the WO and WA configurations yielded a p-value of 0.0039, indicating a statistically significant improvement when an additional sensor was placed on the ankle. Similarly, the difference between WO and W18 also showed statistical significance, with a p-value of 0.00195. This suggests that incorporating multiple sensors across different body locations significantly enhances classification performance compared to using a single wrist-mounted accelerometer.

The comparison between WA and W18 resulted in a p-value of 0.0264. While this difference is statistically significant at the original alpha level of 0.05, the p-value exceeds the adjusted significance threshold after applying Bonferroni correction. The p-value of 0.0264 exceeds the corrected alpha level, meaning that the difference between WA and W18 is not statistically significant after correction. This indicates that, while adding the chest sensor in W18 provides some additional improvement over WA, the enhancement is less substantial compared to the improvements observed when moving from WO to WA or WO to W18.

\section{Discussion}
We evaluated the performance of various neural network architectures with different sensor placement configurations for PAL classification. The CNN+LSTM model achieved the highest average accuracy and F1-score across all configurations, leveraging its ability to combine spatial feature extraction from CNN layers with the temporal modeling of LSTM layers. This combination enables the model to capture both motion patterns and activity progression, outperforming CNN alone. However, the current models were trained on data from healthy participants, so future work should explore their application in patients with OA.

Regarding sensor placement, the WO configuration demonstrated moderate performance, effectively classifies low and medium activities but struggles with high-intensity activities. This limitation is likely due to the inability of a single wrist-mounted accelerometer to fully capture the dynamic motion patterns associated with high-intensity activities. Movements such as jumping involve significant contributions from the lower body, which the wrist sensor alone may not detect effectively.

The addition of an ankle sensor in the WA configuration significantly improved performance, particularly for high-intensity activities. This improvement is likely due to the role of the ankle in human movement kinematics. The ankle experiences direct and pronounced motion changes during high-intensity activities. This additional motion information helps resolve misclassification that arise in the WO configuration.

The WA configuration achieved an accuracy of 92.12\% and an F1-score of 0.90, demonstrating the value of combining multiple sensor locations. This configuration outperformed the WO setup in terms of accuracy, highlighting the advantage of incorporating additional sensors for more precise activity classification. However, the WO configuration remains a practical option for real-world applications due to its simplicity, unobtrusiveness, and ease of use. For individuals with OA, multi-sensor configurations like WA may be particularly beneficial. The ankle sensor can provide detailed gait analysis for OA symptom assessment in daily life and help in designing personalized physical activity programs by offering objective motion data to balance activity levels and prevent overexertion, which can worsen symptoms. Therefore, integrating additional sensors at strategic locations such as the ankle or chest could provide a more comprehensive representation of whole-body movement. Advanced sensor fusion techniques or deep learning models that extract fine-grained motion features from a single wrist sensor could further enhance classification accuracy, particularly for high-intensity activities, without increasing sensor complexity.

Despite these strengths, several limitations should be considered. The relatively small dataset used in this study increases the risk of overfitting, and expanding the dataset with more participants would improve performance and generalization.  

\section{CONCLUSION}

The wrist-only configuration was established as a practical baseline model. However, adding one sensor, particularly at the ankle, improved prediction accuracy and stability. Given the ankle's proximity to the knee, it provides critical motion data that enhances AL classification, especially for applications like knee OA management. These findings highlight the importance of leveraging advanced NN architectures and strategically placed sensors for PAL classification. Future research should validate these findings with larger datasets, incorporate multimodal sensors, and refine models to further advance activity classification and its clinical applications for knee OA.


\begin{thebibliography}{99}

\bibitem{c1} Kraus VB, Sprow K, Powell KE, Buchner D, Bloodgood B, Piercy K, George SM, Kraus WE; 2018 PHYSICAL ACTIVITY GUIDELINES ADVISORY COMMITTEE*. Effects of Physical Activity in Knee and Hip Osteoarthritis: A Systematic Umbrella Review. Med Sci Sports Exerc. 2019 Jun;51(6):1324-1339. doi: 10.1249/MSS.0000000000001944. PMID: 31095089; PMCID: PMC6527143.
\bibitem{c2} Nicholas J Burrows, Benjamin K Barry, Daina L Sturnieks, John Booth, Matthew D Jones, The Relationship Between Daily Physical Activity and Pain in Individuals with Knee Osteoarthritis, Pain Medicine, Volume 21, Issue 10, October 2020, Pages 2481–2495, https://doi.org/10.1093/pm/pnaa096
\bibitem{c3} Mathias Steinach, Hanns-Christian Gunga, Chapter 3 - Exercise Physiology, Editor(s): Hanns-Christian Gunga, Human Physiology in Extreme Environments, Academic Press, 2015, Pages 77-116, ISBN 9780123869470, https://doi.org/10.1016/B978-0-12-386947-0.00003-4
\bibitem{c4} Herrmann SD, Willis EA, Ainsworth BE, Barreira TV, Hastert M, Kracht CL, Schuna Jr. JM, Cai Z, Quan M, Tudor-Locke C, Whitt-Glover MC, Jacobs DR. 2024 Adult Compendium of Physical Activities: A third update of the energy costs of human activities. Journal of Sport and Health Science, 2024;13(1): 6-12.
\bibitem{c5} A. Reiss and D. Stricker. Introducing a New Benchmarked Dataset for Activity Monitoring. The 16th IEEE International Symposium on Wearable Computers (ISWC), 2012.
\bibitem{c6} K. He and X. Zhang and S. Ren and J. Sun. Deep Residual Learning for Image Recognition, CoRR, 1512.03385, 2015, https://doi.org/10.48550/arXiv.1512.03385
\bibitem{c7} Lazaridou A, Martel MO, Cornelius M, Franceschelli O, Campbell C, Smith M, Haythornthwaite JA, Wright JR, Edwards RR. The Association Between Daily Physical Activity and Pain Among Patients with Knee Osteoarthritis: The Moderating Role of Pain Catastrophizing. Pain Med. 2019 May 1;20(5):916-924. doi: 10.1093/pm/pny129. PMID: 30016486; PMCID: PMC6497093.
\bibitem{c8} Nweke, H. F., Teh, Y. W., Al-Garadi, M. A., and Alo, U. R. (2018). Deep learning algorithms for human activity recognition using mobile and wearable sensor networks: State of the art and research challenges. Expert Systems with Applications, 105, 233-261.
\bibitem{c9} Skjødt M, Tully MA, Tsai LT, Gejl KD, Ørtenblad N, Jensen K, Koster A, Visser M, Andersen MS, Caserotti P. Need to Revise Classification of Physical Activity Intensity in Older Adults? The Use of Estimated METs, Measured METs, and VO2 Reserve. J Gerontol A Biol Sci Med Sci. 2024 Jul 1;79(7):glae120. doi: 10.1093/gerona/glae120. PMID: 38703071; PMCID: PMC11215700.
\bibitem{c10}Matthews C.E., Ainsworth B.E., Hanby C., Pate R.R., Addy C., Freedson P.S., Jones D.A., Macera C.A. Development and testing of a short physical activity recall questionnaire. Med. Sci. Sports Exerc. 2005;37:986–994.
\bibitem{c11}Ishikawa-Takata K., Naito Y., Tanaka S., Ebine N., Tabata I. Use of doubly labeled water to validate a physical activity questionnaire developed for the Japanese population. J. Epidemiol. 2011;21:114–121. doi: 10.2188/jea.JE20100079. 
\bibitem{c12}Adams S.A., Matthews C.E., Ebbeling C.B., Moore C.G., Cunningham J.E., Fulton J., Hebert J.R. The effect of social desirability and social approval on self-reports of physical activity. Am. J. Epidemiol. 2005;161:389–398. doi: 10.1093/aje/kwi054. 
\bibitem{c13}Seale J.L., Klein G., Friedmann J., Jensen G.L., Mitchell D.C., Smiciklas-Wright H. Energy expenditure measured by doubly labeled water, activity recall, and diet records in the rural elderly. Nutr. Burbank Los Angel. Cty. Calif. 2002;18:568–573. doi: 10.1016/S0899-9007(02)00804-3. 


\end{thebibliography}
\end{document}